\documentclass[10pt]{article}

\usepackage{amsmath}
\usepackage{latexsym}
\usepackage{amsfonts}
\usepackage{bbm}
\usepackage{graphicx}

\title{Relational Semantics for Databases
         and Predicate Calculus\thanks{
            Report number DCS-343-IR,
            Department of Computer Science,
            University of Victoria, Canada.
        }
      }

\date{}

\author{Philip Kelly\thanks{
        Department of Computer Science,
        University of Victoria, Canada.
        }\\
        \and M.H. van Emden\thanks{
        Department of Computer Science,
        University of Victoria, Canada.
        }
       }

\begin{document}

\maketitle

\abstract{
The relational data model requires a theory of relations
in which tuples are not only many-sorted, but can also
have indexes that are not necessarily numerical.
In this paper we develop such a theory
and define operations on relations that are adequate
for database use.
The operations are similar to those of Codd's
relational algebra, but differ in being based on a mathematically
adequate theory of relations.
The semantics of predicate calculus,
being oriented toward the concept of satisfiability,
is not suitable for relational databases.
We develop an alternative semantics that assigns
relations as meaning to formulas with free variables.
This semantics makes the classical predicate calculus
suitable as a query language for relational databases.
}

\hyphenation{meta-lang-uage}
\hyphenation{data-base data-bases}

\newcommand{\cpp}{\hbox{{\tt C++}}}

\newtheorem{theorem}{Theorem}{}
\newtheorem{definition}{Definition}{}
\newtheorem{lemma}{Lemma}{}
\newtheorem{example}{Example}{}

\newcommand{\D}{\ensuremath{\textbf{D}}}
\newcommand{\F}{\ensuremath{\mathcal{F}}}

\renewcommand{\to}{\rightarrow}
\newcommand{\lra}{\Leftrightarrow}
\newcommand{\set}[1]{\{#1\}}

\newcommand{\Nat}{\ensuremath{\mathcal{N}}} 
\newcommand{\Int}{\ensuremath{\mathcal{I}}} 
\newcommand{\Pred}{\ensuremath{\mathcal{P}}}
\newcommand{\Rat}{\ensuremath{\mathcal{Q}}} 
\newcommand{\Rea}{\ensuremath{\mathcal{R}}} 
\newcommand{\ExtRe}{\ensuremath{\mathcal{R}^{++}}} 
\newcommand{\Flpt}{\ensuremath{\mathcal{F}}} 

\newcommand{\Alp}{\ensuremath{\mbox{\textbf{A}}}} 
\newcommand{\AlpStar} {\ensuremath{\Alp ^\ast}}

\newcommand{\A}{\ensuremath{\mathcal{A}}} 
\newcommand{\I}{\ensuremath{\mathcal{I}}} 
\newcommand{\Pwst}{\ensuremath{\mathcal{P}}} 
\newcommand{\T}{\ensuremath{\mathcal{T}}} 
\newcommand{\M}{\ensuremath{\mathcal{M}}} 
\newcommand{\MI}{\ensuremath{\mathcal{M}_I}}
\newcommand{\cart}{\ensuremath{\mbox{\textsc{cart}}}} 
\newcommand{\apl}{\ensuremath{\mbox{\textsc{apl}}}} 

\newcommand{\pair}[2]{\ensuremath{\langle #1,#2 \rangle}}
\newcommand{\triple}[3]{\ensuremath{[ #1,#2,#3 ]}}
\newcommand{\vc}[2]{\ensuremath{#1_0,\ldots,#1_{#2-1}}}

\newcommand{\id}{\ensuremath{\mbox{id}}}
\newcommand{\argt}{\ensuremath{\mbox{argt}}}
\newcommand{\prd}{\ensuremath{\mbox{prod}}}
\newcommand{\sq}{\ensuremath{\mbox{sq}}}
\newcommand{\tr}{\ensuremath{\triangleright}}

\newcommand{\iot}[1]{\ensuremath{#1}}

\newcommand{\cities}{\ensuremath{\mbox{\textsc{cities}} }}
\newcommand{\parts}{\ensuremath{\mbox{\textsc{parts}} }}
\newcommand{\projects}{\ensuremath{\mbox{\textsc{projects}} }}

\newcommand{\sid}{\ensuremath{\mbox{\textsc{sid}} }}
\newcommand{\sname}{\ensuremath{\mbox{\textsc{sname}} }}
\newcommand{\city}{\ensuremath{\mbox{\textsc{city}} }}
\newcommand{\pid}{\ensuremath{\mbox{\textsc{pid}} }}
\newcommand{\pname}{\ensuremath{\mbox{\textsc{pname}} }}
\newcommand{\pqty}{\ensuremath{\mbox{\textsc{pqty}} }}
\newcommand{\LEQ}{\ensuremath{\mbox{\textsc{leq}} }}

\newcommand{\suppliers}{\ensuremath{\mbox{\textsc{suppliers}} }}

\newcommand{\lee}{\ensuremath{\mbox{\textsc{lee}} }}
\newcommand{\tulsa}{\ensuremath{\mbox{\textsc{tulsa}} }}
\newcommand{\poe}{\ensuremath{\mbox{\textsc{poe}} }}
\newcommand{\ray}{\ensuremath{\mbox{\textsc{ray}} }}
\newcommand{\taos}{\ensuremath{\mbox{\textsc{taos}} }}
\newcommand{\hose}{\ensuremath{\mbox{\textsc{hose}} }}
\newcommand{\tube}{\ensuremath{\mbox{\textsc{tube}} }}
\newcommand{\shim}{\ensuremath{\mbox{\textsc{shim}} }}
\newcommand{\sql}{\ensuremath{\mbox{\textsc{sql}} }}
\newcommand{\etr}{\ensuremath{\mbox{\textsc{etr}} }}

\newcommand{\rid}{\ensuremath{\mbox{\textsc{rid}} }}
\newcommand{\rqty}{\ensuremath{\mbox{\textsc{rqty}} }}
\newcommand{\where}{\ensuremath{\mbox{\textsc{where}} }}
\newcommand{\pc}{\ensuremath{\textsc{pc} }}

\newcommand{\parent}{\ensuremath{\textsc{parent} }}
\newcommand{\child}{\ensuremath{\textsc{child} }}
\newcommand{\mary}{\ensuremath{\textsc{mary} }}
\newcommand{\john}{\ensuremath{\textsc{john} }}
\newcommand{\alan}{\ensuremath{\textsc{alan} }}
\newcommand{\joan}{\ensuremath{\textsc{joan} }}

\newcommand{\answer}{\ensuremath{\textsc{answer} }}

\newcommand{\Sym}{\ensuremath{X}}
\newcommand{\Bool}{\ensuremath{\mathcal{B}}}

\section{Introduction}

Relational databases started with Codd's idea
of queries in a declarative language
to be translated to operations on data
that were justified by a machine-independent semantics.
Such a utopian vision was actually not unrealistic:
predicate calculus is a declarative language
that appeals to a user's intuition
and this language has a machine-independent
semantics in terms of an implementable data-structure,
namely, relations.

Somehow this promising beginning led to \sql.
There were probably several things that went wrong along the way.
In this paper we address one of these.
In analyzing what went wrong one should first acknowledge
the fundamental rightness of Codd's idea.
Next one should examine whether Codd
had a sufficient understanding of what mathematics
has to say about relations and about algebra.
In this paper we address ourselves to the first part: relations.

What Codd found in mathematics was that an $n$-ary relation
is a subset of a Cartesian product
$S_0 \times \cdots \times S_{n-1}$.
It may well be true that the required more general version
is nowhere to be found in mathematics.
If this is the case,
then it is because \emph{for the purposes of mathematicians},
nothing more is needed.
We need to realize that the purposes of database theory
are not necessarily those of mathematicians.
This does not mean that,
as the descent into \sql\ might suggest,
mathematics itself is inadequate,
only that within mathematics developments are needed
that have been overlooked so far.
What needs to be done in database theory
is to use \emph{mathematical methods} to define relations in such a way
as to suit the needs of our theory,
irrespective of whether such a definition
has been sanctioned by prior use in mathematics.
This is the purpose of the present paper.

According to the relational database model,
data are organized by means of relations.
This is a promising point of departure,
provided relations are suitably defined,
as just discussed.
Another doctrine of the relational database model
is that the querying of a database should be based
on an algebra in the mathematical sense.
As far as the present paper is concerned,
this is a premature commitment.
Once a suitable notion of relation is defined,
it is natural to enquire what \emph{operations} on relations
can yield the relations
that are answers to queries.
As we show in this paper,
such operations can be defined set-theoretically,
without any algebra in the mathematical sense
and without commitment to a language in which to express queries.

Although the operations we define
are not designed to constitute an algebra
in the mathematical sense,
they do fit together in a natural way.
This is proved by the fact they allow
a natural translation,
in the sense of compositional semantics,
from predicate calculus
queries and definitions to relational expressions.
Our operations include projection and join,
but defined in a new way to fit our new definition of relation.
They also include a new operation,
which we call \emph{filtering}.
This natural family needs a name;
we propose Elementary Theory of Relations (\etr).
The ``elementary'' is needed
because of its elementary nature
and to distinguish it
from such non-elementary mathematics
as found in works such as \cite{frss00}.

\paragraph{Overview of this paper}

Relations are made up of tuples, and tuples are,
in their full generality, \emph{functions}.
Thus to get relations right,
we need to get tuples right.
To get tuples right,
we do as much as possible with functions.
This gets us all the way up to Cartesian products.
The needed review of terminology and notation for functions
is in Section~\ref{sec:notTerm}.
In Section~\ref{sec:relCodd}
we review Codd's original formulation of the relational data model.
The deficiencies noted here motivate our definition of relations
given in Section~\ref{sec:relMath}.
This serves as basis for what we consider the rational reconstruction
of the relational data model presented in Section~\ref{sec:recon}.
The operations for querying
are worked out in Sections~\ref{sec:relOp} and \ref{sec:queries}.
In Section~\ref{sec:logSem} we show how the operations
take part in the semantics of predicate calculus.
In this way predicate calculus becomes,
once again,
a strong candidate for a database query language.
In Section~\ref{sec:fuWork} we outline what needs to be done.

\section{Set-theoretical terminology}
\label{sec:notTerm}

Loosely speaking, a relation is a set of tuples.
Of course the tuples in such a set have to have something common
if they are to constitute a valid relation.
What precisely they have to have in common
comes down to the precise notion of ``tuple''.
To get this right, we use the definition of Halmos \cite{hlm60},
which views a tuple (``family'' in his terminology) as a function.
In this section we show which of the various definitions
and notations for functions and tuples we use.

$\Nat$ is the set of natural numbers;
$\iota(n)$ is the set
$\{0, \ldots, n-1\}$
of the first $n$ natural numbers.
We write $\subset$ for the subset relation between sets.
We write
\pair{a}{b}
for the ordered pair with elements $a$ and $b$.

\begin{definition}[function]
A \emph{function} is a triple
consisting of a set that is its \emph{source},
a non-empty set that is its \emph{target},
and a \emph{mapping} that associates
with every element of the source
an element of the target\footnote{
Most authors use ``domain'' for ``source'' and
``co-domain'' for ``target''.
We avoid ``domain'' because of its other uses in computer science.
}.
If $s \in S$ is associated by the map with $t \in T$,
then we can write $s \mapsto t$.

The set of all functions with source $S$ and target $T$
is denoted $S \rightarrow T$. This set is often referred to as the
\emph{type} of the functions belonging to it.
Thus we write $f \in S \rightarrow T$ rather
than the usual $f: S \rightarrow T$.

Let $f$ be a function in $S\rightarrow T$ and let $S'$ be a subset
of $S$.
$f(S')$ is defined to be $\{f(x) \mid x \in S'\}$.
In particular, $f(S)$ is the set of elements of $T$ that are a value
of $f$.
\end{definition}

\begin{example} \label{ex:tableCards}
Let $c \in \{\clubsuit, \diamondsuit, \heartsuit, \spadesuit\}
\rightarrow \{black, white, red\}$
be a function with a mapping such that 
$\clubsuit \mapsto \mbox{black}$,
$\diamondsuit \mapsto \mbox{red}$,
$\heartsuit \mapsto \mbox{red}$, and
$\spadesuit \mapsto \mbox{black}$.
One may write the mapping of $c$ more compactly as
\begin{center}
$c = \begin{tabular}{l|l|l|l}
$\spadesuit$ & $\diamondsuit$ & $\heartsuit$ & $\clubsuit$\\
\hline
\mbox{black} & \mbox{red} & \mbox{red} & \mbox{black}
\end{tabular}$,
\end{center}
where the order of the columns is immaterial:
only the pairing of the source and target elements matters.
\end{example}

\begin{example}
For finite $S$ and non-empty finite $T$ we have that
\begin{equation}\label{eq:setPower}
|S \rightarrow T| = |T|^{|S|}
\end{equation}
where $|X|$ is the number of elements in finite set $X$.
Note that $|\{\} \rightarrow T| = 1$,
as there is one function of type $\{\} \rightarrow T$
for any non-empty finite $T$.
Equation (\ref{eq:setPower}) can be extended to cardinal arithmetic
for infinite $S$ or $T$.
\end{example}

\begin{definition}[restriction]
Let $f$ be a function in $S\rightarrow T$ and let $S'$ be a set.
$f \downarrow S'$ is \emph{the restriction of} $f$ \emph{to} $S'$.
It has $S \cap S'$ as source, $T$ as target, and its mapping
is $ x \mapsto f(x)$.
\end{definition}

\begin{definition}[insertion]
If $S' \subset S$ then $i \in S' \rightarrow S$
is \emph{the insertion of} $S'$ \emph{with respect to} $S$
if it has the map $x \mapsto x$.
\end{definition}

\begin{definition}[function sum]
Functions
$f_0 \in S_0 \rightarrow T_0$
and
$f_1 \in S_1 \rightarrow T_1$
are \emph{summable} iff for all $x \in S_0 \cap S_1$, if any,
we have $f_0(x) = f_1(x)$.
If this is the case, then $f_0+f_1$,
the \emph{sum of} $f_0$ \emph{and} $f_1$,
is the function in $(S_0 \cup S_1) \rightarrow (T_0 \cup T_1)$
with map
$x \mapsto f_0(x)$ if $x \in S_0$
and
$x \mapsto f_1(x)$ if $x \in S_1$.
\end{definition}

\begin{example}
Let
$f_0 \in \{a,b\}\rightarrow\{0,1\}$
such that $a \mapsto 0$ and $b \mapsto 1$.
Let
$f_1 \in \{b,c\}\rightarrow\{0,1\}$
such that $b \mapsto 1$ and $c \mapsto 0$.
Then $f_0$ and $f_1$ are summable and
$(f_0+f_1) \in \{a,b,c\}\rightarrow\{0,1\}$
such that $a \mapsto 0$,
$b \mapsto 1$, and
$c \mapsto 0$.
\end{example}

\begin{definition}[function composition]
\label{compDef}
Let $f \in S \rightarrow T$ and $g \in T' \rightarrow U$
with $T' \subset T$.
The \emph{composition} $g \circ f$ \emph{of} $f$ \emph{and} $g$
is the function in $S \rightarrow U$
that has as map $x \mapsto g(f(x))$.
\end{definition}

\begin{figure}[htbp]
\begin{center}
\includegraphics[scale = 0.65]{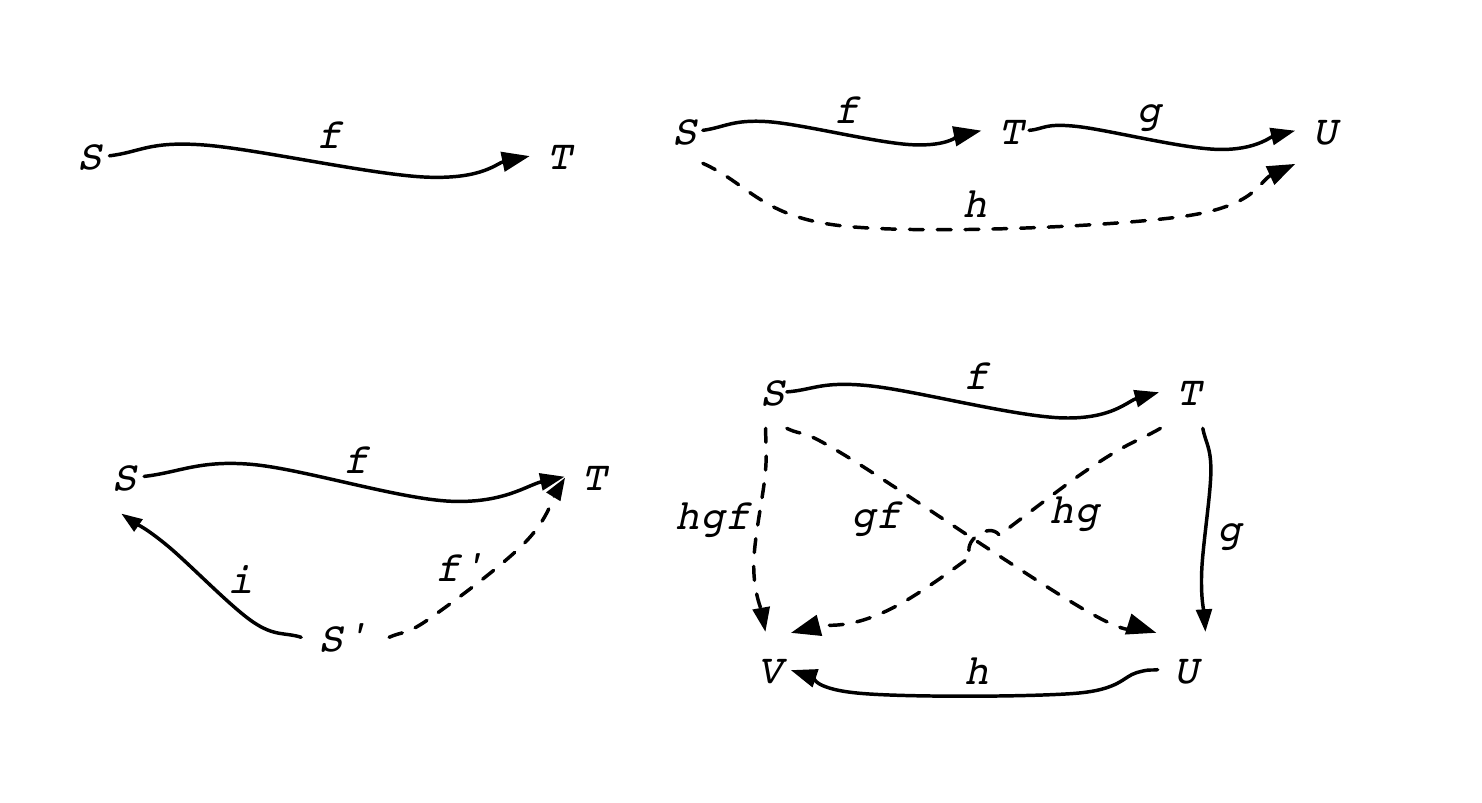}
\end{center}
\caption{\label{fig:fourExamples}
Top left: $f \in S \rightarrow T$.
Top right: $h = g \circ f$.
Bottom left:
$S' \subset S$ and $i \in S' \rightarrow S$
is the insertion function of $S'$ with respect to $S$.
The diagram asserts
$f' = f \circ i$.
Bottom right:
$\mbox{\it gf} = g\circ f$,
$\mbox{\it hg} = h\circ g$, and
$\mbox{\it hgf}= h \circ (g \circ f) = (h\circ g) \circ f$.
}
\end{figure}

For examples, see Figure~\ref{fig:fourExamples}.

If $T$ is a set of disjoint sets,
then there is, for each element $x$ of $\cup T$,
a unique element of $T$ to which $x$ belongs.
This observation suggests the following definition.
\begin{definition}[sorting function]
\label{def:sorting}
Let $T$ be a set of disjoint sets.
The \emph{sorting function} $\sigma_T$ \emph{of} $T$
is the function in $\cup T \rightarrow T$
that maps each element $x$ of $\cup T$ to
the unique set in $T$ to which $x$ belongs.
\end{definition}
The subscript in $\sigma_T$ is often dropped,
hopefully only when the corresponding set of disjoint sets
is clear from the context.

\begin{definition}
If $f \in S \rightarrow T$ is such that
$x \neq y$ implies $f(x) \neq f(y)$,
then $f$ is said to be \emph{injective}.
If $f(S) = T$,
then $f$ is said to be \emph{surjective}.
If $f$ is injective and surjective,
then it is said to be \emph{bijective}.
\end{definition}

\begin{lemma}
If $f \in S \rightarrow T$ is bijective,
then a function $g \in T \rightarrow S$ exists
with map $t \mapsto s$ iff $t = f(s)$.
$g$ is bijective
and is called the \emph{inverse} $f^{-1}$ of $f$. 
\end{lemma}
\subsection{Tuples}

A tuple is a function.
The only thing that is special about tuples
is their terminology.
Let $t \in I \rightarrow T$ be a tuple (i.e. a function)
that maps every $i \in I$ to $t(i) \in T$.
The source is called ``index set''.
$t(i)$ is called the tuple's \emph{component} \emph{indexed} by $i$;
$t(i)$ is often written as $t_i$.

\begin{definition}[empty tuple]
A tuple of type $I \rightarrow T$ is empty
when $I$ is empty.
\end{definition}
For every non-empty $T$ there is a unique empty tuple
of type $\{\} \rightarrow T$.

\begin{definition}[subtuple, sequence]
\label{def:subtuple}
If $t$ is a tuple with index set $I$ and if $I'$ is a subset of $I$,
then $t\downarrow I'$ is the \emph{subtuple} of $t$ defined by $I'$.
If $I = \iota(n)$ for some natural number $n$, then $t$ is called a
\emph{sequence} and $t \downarrow I'$
is a \emph{sub-sequence} of $t$.
\end{definition}

The indexes of a tuple need not be numbers.
Even if they are, they need not be contiguous numbers.
Even if they are,
the least of them need not be 0.
But if the index set of a tuple $t$ is $\iota(n)$ for some natural number $n$, 
then we have the benefit of a concise notation:
$t=[ t_0, \ldots, t_{n-1} ]$\footnote{
We reserve the angle brackets $\langle$ and $\rangle$
for pairs that are not necessarily related to sequences
with $\iota(2)$ as index set.
}.

Consider tuples in $\iota(3) \rightarrow \iota(3)$.
Then $[ 2,0,1 ]$ is a bijection,
so that $[ 2,0,1 ]^{-1}$ exists.
In fact, we have
$
[ 2,0,1 ]^{-1} =
[ 1,2,0 ]
$.

\begin{example}
$t = [ d,b,a,c ]$ implies
that the source of $t$ is $\iota(4)$, and that
$t_0 = d$,
$t_1 = b$,
$t_2 = a$, and
$t_3 = c$.

We can say the following about $t' = t \downarrow \{2,3\}$:
$t'_2 = a$,
$t'_3 = c$,
while
$t'_0$
and
$t'_1$
are not defined.

Note that $t'$, though a subtuple, is not a sequence.
\end{example}

We often consider tuples of which the elements
have different types, that is, belong to different sets.
The typing of such tuples
can be succinctly characterized as follows.

\begin{figure}[htbp]
\begin{center}
\includegraphics[scale = 0.85]{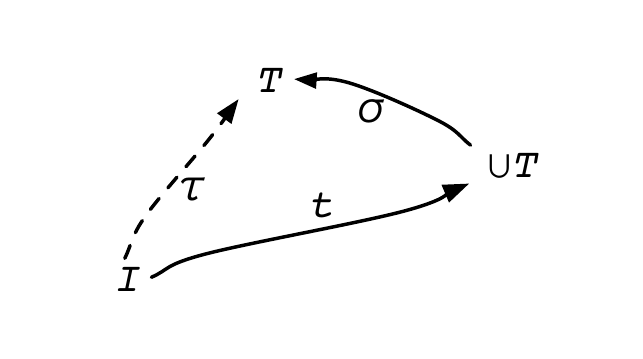}
\end{center}
\caption{\label{fig:sig}
$\tau \in I \rightarrow T$ is the signature
of $t \in I \rightarrow \cup T$ because $\tau = \sigma \circ t$.
}
\end{figure}

\begin{definition}[signature, sorting of tuples]
Let $I$ be a set of indexes, $T$ a set of disjoint sets,
$\tau$ a tuple in $I \rightarrow T$, and
$t$ a tuple in $I \rightarrow \cup T$.
We say that $t$ is \emph{sorted by} 
(or \emph{has signature}) $\tau$ 
iff
$\tau = \sigma \circ t$,
where $\sigma$ is the sorting function of $T$.
\end{definition}
That is,
$t$ is sorted by $\tau$
if we have $t(i) \in \tau(i)$ for all $i \in I$:
the type of each element of $t$ is determined by $\tau$. 
See Figure~\ref{fig:sig}.

Let $\Bool = \{\textsf{t},\textsf{f}\,\}$.
Let $T = \{\Bool,\Nat\,\}$ and $\tau = [ \Bool,\Nat,\Nat\, ]$.
Then
$[ \textsf{t},4,5 ]$ has signature $\tau$,
while
$[ \textsf{t},\textsf{t},5 ]$
does not.

\subsection{Pattern matching}

We want to express mathematically
a certain way in which one tuple ``matches'' another,
like $\triple{a}{a}{c}$ matching $\triple{x}{x}{z}$
because we can transform
$\triple{x}{x}{z}$
to
$\triple{a}{a}{c}$
by substituting $a$ for $x$ and $c$ for $z$
in $\triple{x}{x}{z}$.

For example $\triple{a}{a}{a}$
matches $\triple{x}{x}{z}$,
but not the other way around.
While $\triple{a}{a}{c}$
does match $\triple{x}{x}{z}$,
$\triple{a}{b}{c}$ does not.
Also, both
$\triple{a}{a}{c}$
and
$\triple{a}{b}{c}$
match
$\triple{x}{y}{z}$.
It will be useful for later to note
that $\triple{x}{y}{z}$ matches
\emph{every} triple in
$\triple{0}{1}{2} \rightarrow \{a,b,c\}$. 

We can make these intuitions precise in the following way.
\begin{definition}[compatible, matching, pattern, substitution]
\label{def:matching}
Let $T$ be a set of disjoint sets.
Let $t$ be a tuple typed by $\tau \in I \rightarrow T$
and let $p \in I \rightarrow X$ be a tuple typed by  
$\varphi \in \Sym \rightarrow T$.
The typing of $p$ is \emph{compatible with} $\tau$ in the sense that
$\tau = \varphi \circ p$.
We define a binary relation between $t$ and $p$ by the condition
that there exists an $s \in \Sym \rightarrow \cup T$
such that $t = s \circ p$.
In this context we say that $t$ \emph{matches pattern} $p$
\emph{with matching substitution} $s$.
\end{definition}

\begin{figure}[htbp]
\begin{center}
\includegraphics[scale = 0.85]{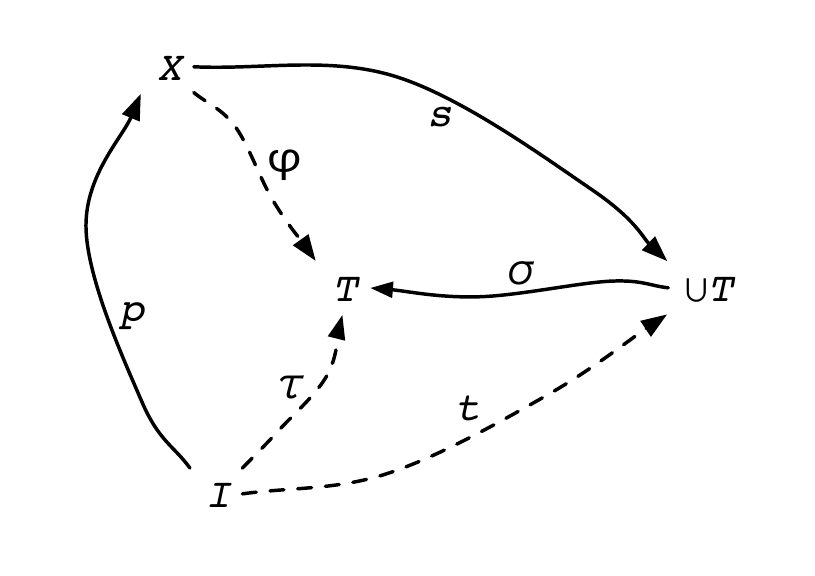}
\end{center}
\caption{
\label{fig:network} 
To Definition~\ref{def:matching}.
}
\end{figure}

Let us see how this works out for one of the examples just mentioned.
We have $I =\{0,1,2\}$,
$T=\{\{a,b,c\}\}$,
$\Sym = \{x,y,z\}$.
Thus 
$\triple{a}{a}{c} \in I \rightarrow \cup T$
matches
$\triple{x}{x}{z} \in I \rightarrow \Sym$
because there exists $s \in \Sym \rightarrow \cup T$
such that 
$\triple{a}{a}{c} = s \circ \triple{x}{x}{z}$.
Several such values for $s$ exist:
$s = \begin{tabular}{l|l|l}
$x$ & $y$ & $z$ \\
\hline
$a$ & $a$ & $c$
\end{tabular}$,
$s = \begin{tabular}{l|l|l}
$x$ & $y$ & $z$ \\
\hline
$a$ & $b$ & $c$
\end{tabular}$, and
$s = \begin{tabular}{l|l|l}
$x$ & $y$ & $z$ \\
\hline
$a$ & $c$ & $c$
\end{tabular}$.
But no $s \in \Sym \rightarrow \cup T$ exists such that
$\triple{a}{b}{c} = s \circ \triple{x}{x}{z}$.
So $\triple{a}{b}{c}$ does not match pattern $\triple{x}{x}{z}$.

In Figure~\ref{fig:network} we have $t = s \circ p$:
$t$ matches $p$ with substitution $s$.
Are there patterns $p$ that match every tuple $t$?
In fact, if $p$ is a bijection, then $p$ (which is a function)
has an inverse $p^{-1}$ and we have
$t \circ p^{-1} = s \circ p \circ p^{-1} = s.$
This gives the matching substitution with which
an arbitrary $t$ matches $p$.

\subsection{Cartesian products}

In mathematics a Cartesian product is often a set of tuples
$[ x_0, \ldots, x_{n-1}]$
where the elements of the tuples belong to a set $D$.
Such a Cartesian product is denoted $D^n$.
Expressed more formally,
$D^n$ is the set of tuples
that have in common the signature $\tau \in I \rightarrow T$,
where $I = \iota(n)$ and $T= \{D\}$.
In mathematics one sometimes generalizes the notion of Cartesian product
to $D_0 \times \cdots \times D_{n-1}$,
which is the set of tuples with the common signature
$\tau \in I \rightarrow T$,
where $I = \iota(n)$ and $T= \{D_0, \ldots, D_{n-1}\}$
and $i \mapsto D_i$ for all $i \in \iota(n)$.

The two examples suggest a further generalization:
to allow the index set $I$ to be an arbitrary set.
This generalization is rarely, if ever, needed in mathematics,
but is useful in data management\footnote{
Often referred to as ``data modeling'',
which is suspect: what one models is the \emph{world};
data is what the model is made of.
}.

\begin{definition}[Cartesian product]
\label{def:cartProd}
Let $I$ be a set of indexes, $T$ a set of disjoint sets, and
$\tau$ a tuple in $I \rightarrow T$.
The \emph{Cartesian product on} $\tau$,
denoted $\cart(\tau)$, is the set
of all tuples that have signature $\tau$.
\end{definition}

\section{The relational model according to Codd}
\label{sec:relCodd}
We have now all the ingredients
for an adequate definition of ``relation''.
Before proceeding to such a definition
we review in this section
Codd's original one,
from section 1.3 (``A Relational View of Data'') of \cite{codd70}. 
\begin{quote}
\emph{
The term \emph{relation} is used here in its accepted mathematical sense.
Given sets $S_1,\ldots,S_n$ (not necessarily distinct),
$R$ is a relation on these $n$ sets if it is a set of $n$-tuples
each of which has its first element from $S_1$, its second
element from $S_2$, and so on.
We shall refer to $S_j$ as the $j$th \emph{domain} of $R$. \\
}
\end{quote}

In other words, mathematically speaking,
a relation is a subset of a Cartesian product of the form
$S_1 \times \cdots \times S_n$.
Note that $S_j$, not $j$, is the domain.
Thus $S_i$ and $S_j$ may be the same set,
even though $i \not = j$.
The above quote continues with:

\begin{quote}
\emph{
For expository reasons, we shall frequently make use of an array
representation of relations, but it must be remembered that this 
particular representation is not an essential part of the relational
view being expounded. An array which represents an $n$-ary relation
$R$ has the following properties:
\begin{enumerate}
\item
Each row represents an $n$-tuple of $R$.
\item
The ordering of rows is immaterial.
\item
All rows are distinct.
\item
The ordering of columns is significant --- it corresponds to
the ordering $S_1,\ldots,S_n$ of the domains on which $R$ is defined.
\item
The significance of each column is partially conveyed by labeling
it with the name of the corresponding domain.
\end{enumerate}
}
\end{quote}

Codd gives as example of such an array
the one shown in Figure~\ref{CoddArr1}.
He observes that this example
does not illustrate why the order of the columns matters.
For that he introduces the one in Figure~\ref{CoddArr2}.
\begin{figure}[!htb]
\begin{center}
\rule{10cm}{0.005in}
\begin{tabular}{ccccc}
supply & (supplier & part & project & quantity)  \\
       &  1        & 2    & 5       & 17         \\
       &$\ldots$   &$\ldots$ &$\ldots$ & $\ldots$ \\
\end{tabular}
\caption{
\label{CoddArr1}
Codd's first example: domains all different.
}
\rule{10cm}{0.005in}
\end{center}
\end{figure}

\begin{figure}[!htb]
\begin{center}
\rule{10cm}{0.005in}
\begin{tabular}{ccccc}
& component & (part  & part & quantity)  \\
&           &  1     & 5    & 9         \\
&           &$\ldots$& $\ldots$ &$\ldots$ \\
\end{tabular}
\caption{
\label{CoddArr2}
Codd's second example: domains not all different.
}
\rule{10cm}{0.005in}
\end{center}
\end{figure}

He explains Figure~\ref{CoddArr2} as follows.
\begin{quote}
$\ldots$ two columns may have identical headings (indicating
identical domains) but possess distinct meanings with respect
to the relation.
\end{quote}

We can take it that in Figure~\ref{CoddArr2}
we have $n = 3$, $S_1 = S_2 = \mbox{ part}$,
and $S_3 = \mbox{ quantity}$.
As $S_1, \ldots, S_n$ need not all be different,
columns can only identified by $\{1,\ldots , n\}$.

Codd goes on to point out that in practice $n$
can be as large as thirty
and that users of such a relation
find it difficult to refer to each column
by the correct choice among the integers $1,\ldots,30$.
According to Codd, the solution is as follows.

\begin{quote}
\emph{
Accordingly, we propose that users deal, not with relations,
which are domain-ordered, but with \emph{relationships},
which are their domain-unordered counterparts.
To accomplish this, domains must be uniquely
identifiable at least within any given relation,
without using position.
Thus where there are two or more identical domains,
we require in each case that the domain name be qualified
by a distinctive \emph{role name},
which serves to identify the role played
by that domain in the given relation.
} 
\end{quote}

Apparently
\begin{itemize}
\item
``Role names'' are non-numerical identifications of columns.
\item
A database is a collection of relations and relationships.
The former in case all of $S_1, \ldots, S_n$ are different;
the latter if not.
\item
Relations are defined mathematically as
subsets of $S_1 \times \cdots \times S_n$.
Relationships are something else.
Their mathematical definition is not given.
\end{itemize}

However, later on page 380 of \cite{codd70}, we find
\begin{quote}
\emph{
$\ldots$ it is proposed that most users should interact
with a relational model of the data consisting of a collection
of time-varying relationships (rather than relations).
} 
\end{quote}

That is clear enough: no longer any use for the mathematically
defined relations. But the next sentence is 

\begin{quote}
\emph{
Each user need not know more about any relationship
than its name together with the names of its domains
(role qualified whenever necessary).
} 
\end{quote}

Apparently role qualification is not always deemed necessary.
We interpret this as the proposal to have data in the form of a relation
when the degree is small and the domains are all different.
If the degree is not small or if there are repeated domains,
then the data are supposed to be in the form of a ``relationship''.

We consider such dual treatment of data unsatisfactory.
Our proposal is to have all domains ``role-qualified''
so that all data are in the form of what Codd calls ``relationships''.
Of course relationships are not subsets of
$S_1 \times \cdots \times S_n$.
Codd did not say what they were, mathematically.
We propose to refer to ``relationships'' 
as ``relations'',
to be defined mathematically in the next section.

\section{A mathematical definition of relation}
\label{sec:relMath}

In modeling data,
Codd wanted that
``the term \emph{relation} be used in its accepted mathematical sense''.
The problem with this is that this accepted sense
arose out of typically mathematical concerns.
It is not to be expected
that this sense is adequate
when one shifts the area of inquiry
from pure mathematics to data management.

However, mathematics is ready to help out
once one has decided on a suitable extra-mathematical application,
and the relational representation of data is one such application.
All that is needed is to drop the notion
that the tuples constituting relations
have to be indexed numerically.
In fact, the indexes can be the ``role names''
of which Codd noticed that they would be useful
when there are multiply occurring domains
or when the tuples have many elements.

For this reason we have defined a Cartesian product
to consist of tuples with index sets
that are not necessarily numerical.
This is the needed generalization;
we can, as usual, define a relation
as a subset of a Cartesian product,
but then in the sense of Definition~\ref{def:cartProd}.
As this Cartesian product may have any signature,
that signature must be included in the definition of relation.
Hence:

\begin{definition}[relation]\label{relDef}
A \emph{relation} is a pair $\pair{\tau}{E}$
where $\tau$, the \emph{signature} of the relation,
is a tuple of type $I \rightarrow T$, where $T$ is a set of disjoint sets,
and $E$, the \emph{extent} of the relation,
is a subset of $\cart(\tau)$.
\end{definition}

As all tuples in the extent of a relation have the same signature,
the tabular notation of Example~\ref{ex:tableCards} is convenient:
every tuple can be a row under the same headings.
If the index set has fewer elements than the extent, then
it may be preferable to transpose the table, as is done in
the following example.

\begin{example}
Consider the relation $\pair{\tau}{E}$
where
$$\tau \in \{divisor, dividend, quotient\} \rightarrow \{\iota(6)\}$$
and
$E =\{t\mid t \mbox{ is typed by } \tau \mbox{ and }
  t_{dividend} = t_{divisor}\times t_{quotient}
\}$, that is

$E$ = 
\begin{center}
\begin{tabular}{l||c|c|c|c|c|c|c|c|c|c|c}
dividend&0&1&2&2&3&3&4&4&4&5&5  \\
divisor &0&1&2&1&3&1&4&2&1&5&1  \\
quotient&0&1&1&2&1&3&1&2&4&1&5  \\
\end{tabular}.
\end{center}
\end{example}

\begin{example}
The difficulty with Codd's example in Figure~\ref{CoddArr1}
is that the column headings have to serve both as indexes
and as domains.
Let us adopt Codd's names for the domains.
Let us introduce role names and collect these into the index set 
$I = \{
\mbox{sup},
\mbox{prt},
\mbox{pct},
\mbox{qty}
\}$.

Then we define
$\mbox{supply} = $ \pair{\tau}{E}
as relation suitable to represent Codd's data
with $\tau \in I \rightarrow T$.
Here
$T = \{
\mbox{supplier},
\mbox{part},
\mbox{project},
\mbox{quantity}
\}$,
where we assume that the elements of $T$ are disjoint sets.
The relation is now $\pair{\tau}{E}$, where
\begin{center}
$\tau = $
\begin{tabular}{c|c|c|c}
sup & prt & pct & qty  \\
\hline 
supplier & part & project & quantity \\
\end{tabular},
\end{center}
and
\begin{center}
$E = $
\begin{tabular}{c|c|c|c}
\mbox{sup} &
\mbox{prt} &
\mbox{pct} &
\mbox{qty} \\
\hline 
sup1 & prt2 & pct5 & 17 \\
$\ldots$ & $\ldots$ & $\ldots$ & $\ldots$
\end{tabular}.
\end{center}
\end{example}

\begin{example}

The difficulty with Codd's example in Figure~\ref{CoddArr2}
is that the column headings cannot serve both as indexes
and as domain names because of the repeated domain.
Let us adopt Codd's names for the domains.
Let us introduce role names and collect these into the index set 
$I = \{
\mbox{prt},
\mbox{assembly},
\mbox{qty}
\}$.

Then we define $\mbox{component} =$
\pair{\tau}{E}
as relation suitable to represent Codd's data
with $\tau \in I \rightarrow T$.
Here $T = \{
\mbox{part},
\mbox{quantity}
\}$
where we assume that the elements of $T$ are disjoint sets.
The relation is now $\pair{\tau}{E}$, where
\begin{center}
$\tau = $
\begin{tabular}{c|c|c}
prt & assembly & qty  \\
\hline 
part & part & quantity \\
\end{tabular},
\end{center}
and
\begin{center}
$E = $
\begin{tabular}{c|c|c}
prt & assembly & qty  \\
\hline 
prt1 & assy5 & 9 \\
$\ldots$ & $\ldots$ & $\ldots$
\end{tabular}.
\end{center}
\end{example}

\begin{example}
In a relation \pair{\tau}{E}
we can have that $\tau$ is $\{\} \rightarrow T$.
As there is only one tuple of type $\tau$ (the empty one)
there are only two possibilities for $E$ as subset of
$\{\} \rightarrow T$.
\end{example}

\section{A reconstruction of the relational model
according to set theory}
\label{sec:recon}

Let us now see what the relational format for data
would look like to someone who knows some set theory
and who was only told the general idea of \cite{codd70}
without the details as worked out by Codd.
We first look at how data are stored,
then how they are queried.

\subsection{Using base relations as repositories for data}
Let us start with one way of characterizing a database,
relational or not.
The information in a database
describes various aspects of things
like parts of a gadget, books in a library, and so on. 
The things may be concrete objects
with a certain degree of permanence.
They may also be transient states of affairs,
such as transactions, events,
or employees in a company.
We refer to these various things as \emph{items}.

There is no limit to the information
one can collect about an item as it exists in the world.
Hence one performs an act of abstraction
by deciding on a set of \emph{attributes}
that describe the item
and one determines what is the \emph{value}
of each attribute for this  particular item.
A consequence of this abstraction
is that it cannot distinguish between items
for which all attributes have the same value.
The set of attributes has to be comprehensive enough
that such identification
does not matter for the purpose of the database.

The foregoing is summarized in the first of the following points.
The remaining points constitute a reconstruction
of the main ideas of a relational database as suggested by the first point.

\begin{enumerate}
\item
A database is a description of a world populated by \emph{items}.
For each item, the database lists the
\emph{values} of the applicable \emph{attributes}.
\item
The database presupposes a set of attributes, and for each attribute,
a set of allowable values for this attribute.
Such sets of admissible values are called \emph{domains}.
Let $I$ be the set of attributes and let $T$ be a set of domains,
which we assume to be mutually disjoint.
As each attribute has a uniquely determined domain,
this information is expressed by a function, say $\tau$,
that is of type $I \rightarrow T$.
\item
The description of each item
is an association of a value with each attribute
that is applicable to the item.
If $I' \subset I$ is the set of attributes of the item,
then $I'$ is the index set of the tuple describing it.
The signature of this tuple is $\tau \downarrow I'$.
\item
Let $I_0, \dots, I_{n-1}$
be the index sets (not necessarily different) of the tuples
occurring in the database.
Then
$
\tau\downarrow I_0,
\ldots,
\tau\downarrow I_{n-1}
$
are the signatures occurring in the database.
For all $i$,
let $E_i$ be the set of tuples that are in the database
and have signature $\tau\downarrow I_i$.
Then the database consists of the relations
$$
\pair{\tau\downarrow I_0}{E_0},
\ldots,
\pair{\tau\downarrow I_{n-1}}{E_{n-1}}.
$$
\item
The life cycle of a database includes a design phase
followed by a usage phase.
In the design phase $I$, $T$, and
$\tau \in I \rightarrow T$
is determined,
as well as the subsets
$I_0, \ldots, I_{n-1}$
of $I$.
This is the database \emph{scheme}.
In the usage phase the extents
$E_0, \ldots, E_{n-1}$
are added and modified.
With the extents added, we have a database \emph{instance}.
\end{enumerate}

\subsection{Using computed relations as answers to queries}

We may need rearrangements of the information in the base relations.
This need can be met by using the relation resulting from evaluating
an expression in terms of base relations and relational operations.
We regard such an expression as a \emph{query} and its value
as the \emph{answer} to the query.

\section{Operations on relations}
\label{sec:relOp}

\subsection{Boolean operations}
Among relations of the same signature,
certain operations are defined
that mirror the boolean set operations.
Of these, intersection is useful as an auxiliary to the definition
of relational join; see Definition~\ref{def:join}.
We also list the cognates of some of the most common
set operations.
\begin{definition}\label{def:boolean}
$$
\begin{array}{lcrcl}
\mbox{intersection:} &&
\pair{\tau}{E_1} \cap \pair{\tau}{E_2} &\stackrel{def} {=}& 
              \pair{\tau}{E_1\cap E_2}         \\
\mbox{union:}        &&
\pair{\tau}{E_1} \cup \pair{\tau}{E_2} &\stackrel{def}{=}&
              \pair{\tau}{E_1\cup E_2}         \\
\mbox{difference:}        &&
\pair{\tau}{E_1} \setminus \pair{\tau}{E_2} &\stackrel{def}{=}&
              \pair{\tau}{E_1\setminus E_2}  \\
\mbox{complement:}        &&
\pair{\tau}{E}^C &\stackrel{def}{=}&
              \pair{\tau}{\cart(\tau)\setminus E}
\end{array}
$$
\end{definition}

\subsection{Projection}
If every tuple of a relation is transformed in the same way
into the subtuple (see Definition~\ref{def:subtuple})
defined by a subset of the index set,
then the result is a relation.
\begin{definition}[projection]\label{def:proj}
Let $\tau$ be in $I \rightarrow T$,
where $T$ is a set of disjoint sets
and $I$ is an index set.
Let $J$ be a subset of $I$.
The \emph{projection}
\emph{on} $J$ of the relation
$\pair{\tau}{E}$ is
written $\pi_J(\pair{\tau}{E})$
and is defined to be
the relation $\pair{\tau\downarrow J}{\{t \downarrow J \mid t \in E \}}$.
\end{definition}

\subsection{Filtering}
Suppose we have a relation $r = \pair{\tau}{E}$
with
$E$ a set of tuples of signature $\tau$
where $\tau \in I \rightarrow T$ and $T$ a disjoint set of sets.
Consider now a pattern $X$
that is typed compatibly with the tuples of $r$;
that is, $p \in I \rightarrow \Sym$
and $\varphi \in \Sym \rightarrow T$
such that $\tau = \varphi \circ p$.
Here $\Sym$ is the set containing
the components of the patterns.
We can think of them as variables or, better,
as \emph{indeterminates}.
What matters more than their name is that they
are what is substituted when a pattern $p$ matches a tuple $t$
from $r$. 
The condition for such matching
is the one in Definition~\ref{def:matching}.

In the setting of Figure~\ref{fig:network}
one can consider the set of elements of $E$
that match the pattern $p$.
The corresponding substitutions are a set of tuples with the same
signature, hence can be the extent of a suitably defined relation.
This relation is the result of ``filtering'' $r$ with $p$.
This result is denoted $r:p$.

\begin{definition} [filtering] \label {def:filtering}
Let $T$ be a set of mutually disjoint sets
and let $\sigma \in \cup T \to T$ be the function
that maps an element to the set which it belongs.
Let $\tau \in I \to T$ determine the types associated with
the indexes and let $p \in I \to X$ determine the variables
associated with the indexes.
Let $s \in X \to \cup T$ substitute domain elements for variables
and let $\phi \in X \to T$ determine the types of the variables.
Then the \emph{filtering} $:$ of a relation by $p$ is defined by
\begin{eqnarray*}
\pair{\tau}{E} : p &=& \pair{\varphi}{E'} \mbox{ where } \\
             E'    &=&
   \{s \in \Sym \rightarrow \cup T \mid
         \exists t \in E \mbox{ such that } t = s \circ p\}
\end{eqnarray*}
\end{definition}

The case where $p$ is a bijection is interesting,
as it gives rise to the following equalities:

For all $E \subset \cart(\tau)$ 
$$
(\pair{\tau}{E} : p) : p^{-1} = \pair{\tau}{E}
$$
and, for all $E \subset \cart(\varphi)$ 
$$
(\pair{\varphi}{E} : p^{-1}) : p = \pair{\varphi}{E}.
$$
The relations 
$\pair{\tau}{E}$
and
$\pair{\varphi}{E'}$
can be thought of as renamings of each other,
with $p$ and $p^{-1}$ as renaming schemes.

If $p$ is not a bijection,
then there is potentially more going in $\pair{\tau}{E} : p$
than a renaming,
so it would be a mistake to call it that.
What is it that happens in $\pair{\tau}{E} : p$
with an arbitrary $p$?
We think that ``filtering'' is a good name.

\subsection{Join}
Given a relation $r$, one may be interested in relations
that have $r$ as projection.
There is a largest such, which is the ``cylinder'' on $r$.
This gives the idea;
to get the signatures right, see the following definition.

\begin{definition}[cylinder]
For $i \in \{0,1\}$,
let $\tau_i$ be in $I_i \rightarrow T_i$,
such that $\tau_0$ and $\tau_1$ are summable;
$T_i$ is a set of disjoint sets.
Let $\pair{\tau_0}{E_0}$ be a relation.
The \emph{cylinder}
\emph{with respect to} $\tau_1$
\emph{on} $\pair{\tau_0}{E_0}$
is written as
$\pi^{-1}_{\tau_1}(\pair{\tau_0}{E_0})$
and is defined to be relation
$$\pair{\tau_0+\tau_1}{\{t \in \cart(\tau_0+\tau_1) \mid t\downarrow I_0 \in E_0\}}.$$
\end{definition}
The notation $\pi^{-1}$ suggests some kind of inverse of projection.
It is suggested by facts such as
$\pi_{I_0}(\pi^{-1}_{\tau_1}(\pair{\tau_0}{E_0})) =
\pair{\tau_0}{E_0}$.

\begin{example}
\label{cylEx}
Let
$I_0 = \{a,b,c\}$
and
$I_1 = \{b,c,d\}$.
Let
$\tau_0$ be in $I_0 \rightarrow \{\{0,1\}\}$
and
$\tau_1$ be in $I_1 \rightarrow \{\{0,1\}\}$.
Consider the relations
$\pair{\tau_0}{E_0}$
and
$\pair{\tau_1}{E_1}$,
where
$E_0 = $
\begin{tabular}{c|c|c}
a & b & c  \\
\hline
0 & 0 & 1  \\
1 & 1 & 0  \\
\end{tabular} 
and
$E_1 = $
\begin{tabular}{c|c|c}
b & c & d  \\
\hline
0 & 1 & 0  \\
1 & 1 & 1  \\
\end{tabular}
.
We have
$
\pi^{-1}_{\tau_1}(\pair{\tau_0}{E_0}) =
\pair{\tau_0+\tau_1}{C_0}
$
and
$
\pi^{-1}_{\tau_0}(\pair{\tau_1}{E_1}) =
\pair{\tau_0+\tau_1}{C_1}
$
where
$C_0 = $
\begin{tabular}{c|c|c|c}
a & b & c & d  \\
\hline
0 & 0 & 1 & 0  \\
0 & 0 & 1 & 1  \\
1 & 1 & 0 & 0  \\
1 & 1 & 0 & 1  \\
\end{tabular}
and
$C_1 = $
\begin{tabular}{c|c|c|c}
a & b & c & d  \\
\hline
0 & 0 & 1 & 0  \\
1 & 0 & 1 & 0  \\
0 & 1 & 1 & 1  \\
1 & 1 & 1 & 1  \\
\end{tabular}
.
\end{example}

\begin{definition}[join]\label{def:join}
For $i \in \{0,1\}$ let there be relations $\pair{\tau_i}{E_i}$,
with $\tau_i \in I_i \rightarrow T_i$ and $T_i$ a set of disjoint sets.
If
$\tau_0$
and
$\tau_1$
are summable, then the \emph{join}
of
$\pair{\tau_0}{E_0}$ and $\pair{\tau_1}{E_1}$
is written as
$\pair{\tau_0}{E_0} \Join \pair{\tau_1}{E_1}$
and defined to be
$
\pi^{-1}_{\tau_1}(\pair{\tau_0}{E_0})
\cap
\pi^{-1}_{\tau_1}(\pair{\tau_1}{E_1})
$.
\end{definition}

The intersection in this definition is defined because of the assumed
summability of $\tau_0$ and $\tau_1$.
The signature of $\pair{\tau_0}{E_0} \Join \pair{\tau_1}{E_1}$
is $\tau_0 + \tau_1$.

\begin{example}
Let
$\tau_0$,
$\tau_1$,
$E_0$, and
$E_1$
be as in Example~\ref{cylEx}.
Then we have
$
\pair{\tau_0}{E_0}
\Join
\pair{\tau_1}{E_1}
=
\pair{\tau_0+\tau_1}{E}
$
where
$E = $
\begin{tabular}{c|c|c|c}
a & b & c & d  \\
\hline
0 & 0 & 1 & 0  \\
\end{tabular}
.
\end{example}

\begin{figure}
\begin{center}
\parbox{2in}{
$\suppliers =$\\
\begin{tabular}{l|l|l} 
\sid & \sname & \city     \\
\hline
321  & \lee   & \tulsa    \\   
322  & \poe   & \taos     \\
323  & \ray   & \tulsa    \\
\end{tabular}
}
$\;\;\;$
\parbox{2in}{
$ \parts = $\\
\begin{tabular}{l|l|l|r} 
\pid & \pname & \sid  & \pqty   \\
\hline
213  & \hose   & 322  & 13  \\   
214  & \tube   & 321  & 6   \\
215  & \shim   & 322  & 18  \\
\end{tabular}
}
\end{center}
\begin{center}
\parbox{2in}{
$ \projects = $ \\
\begin{tabular}{l|l|r} 
\rid & \pid & \rqty   \\
\hline
132  & 215  & 2  \\   
133  & 214  & 11   \\
134  & 213  & 18  \\
\end{tabular}
}
\end{center}
\caption{\label{fig:citiesParts}
The relations for Example~\ref{ex:citiesParts}.
}
\end{figure}

\begin{example}\label{ex:citiesParts}
Consider the relations
shown in the tables in Figure~\ref{fig:citiesParts}.
Each table consists of a line of headings
followed by the line entries of the tables.
The line entries represent the tuples of the relations.
Each table has three such lines.

The following abbreviations are used.
In the \suppliers\ table,
\sid\ for supplier ID,
\sname\ for supplier name, and
\city\ for supplier city.
In the \parts\ table,
\pid\ for part ID,
\pname\ for part name, and
\pqty\ for part quantity on hand.
In the \projects\ table,
\rid\ for project ID and
\rqty\ for part quantity required.

We want to know the part names and cities
in which there is a supplier
with a sufficient quantity on hand
for at least one of the projects.
\end{example}

Let the tables in Figure~\ref{fig:citiesParts}
be the relations
$\suppliers = \pair{\tau_0}{E_0}$,
$\parts = \pair{\tau_1}{E_1}$, and
$\projects = \pair{\tau_2}{E_2}$.
In addition, there is a relation in the query for which there is
no table, namely the less-than-or-equal relation.
Mathematically, there is no reason to treat it differently from
the relations stored in tables.
Hence, we also include
it as $\LEQ = \pair{\tau_3}{E_3}$.

The index sets of
$\tau_0$,
$\tau_1$,
$\tau_2$
are the sets of the column headings of the tables
for \suppliers, \parts, and \projects, respectively:
\begin{itemize}
\item
for
$\tau_0$ the index set is $\{\sid, \sname, \city\}$,
\item
for
$\tau_1$ it is $\{\pid, \pname, \sid, \pqty\}$,
\item
for
$\tau_2$ it is $\{\rid, \pid, \rqty\}$,
\item
for
$\tau_3$ it is $\{\rqty, \pqty\}$.
\end{itemize}

The extents
$E_0$,
$E_1$, and
$E_2$
are as described in Figure~\ref{fig:citiesParts}.
Moreover,
$E_3 = \{t \in \cart(\tau_3) \mid t_{\rqty} \leq t_{\pqty}\}$.

\begin{definition}[relational product]\label{def:relProd}
For $i \in \{0,1\}$ let there be relations $\pair{\tau_i}{E_i}$,
with $\tau_i \in I_i \rightarrow T_i$ and $T_i$ a set of disjoint sets.
If $I_0$ and $I_1$ are disjoint,
then the \emph{relational product} of 
$\pair{\tau_0}{E_0}$
and
$\pair{\tau_1}{E_1}$
is defined and it is defined to be equal to
$\pair{\tau_0}{E_0} \Join \pair{\tau_1}{E_1}$.

\end{definition}

The word ``join'' is borrowed from relational databases
\cite{cdd72,bthlvn94},
where ``natural join'' denotes a similar operation.
Definition \ref{def:relProd} shows that an operation
reminiscent of Codd's relational product is a special case of join.

The following lemma consists of assorted equalities
selected to build intuition.

\begin{lemma}\label{lemma:assorted}

\begin{eqnarray*}
\langle \tau, E_0\rangle \Join \langle \tau, E_1\rangle
     &=& 
\langle \tau, E_0 \cap E_1 \rangle 
\\
\pi_{\tau_1}^{-1}(\langle \tau_0, E_0 \rangle)
     &=& 
\langle \tau_0, E_0\rangle  \Join 
\langle \tau_1, \cart(\tau_1)\rangle
\\
\langle \tau_0, E_0 \rangle  \Join 
\langle \tau_1, \cart(\tau_1)\rangle
& = &
(\langle \tau_0, \cart(\tau_0) \rangle  \Join 
 \langle \tau_1, E_1\rangle)
\end{eqnarray*}

\end{lemma}

\section{Queries}
\label{sec:queries}

Certain queries can be expressed as projections of joins.
In such simple queries each relation is referred to only once.
When a query needs to refer to the same relation more than once,
\sql\ uses the renaming operation. 
In this section we show how filtering
can take the place of renaming.

\subsection{Projections of joins as queries}

The \sql\ query in Figure~\ref{fig:SQL}
can be understood as specifying
a relation defined in terms of given relations
understood as in Definition~\ref{relDef}.
In this query, projection and join
can be understood according to
Definitions \ref{def:proj} and \ref{def:join}, respectively.

Consider the expression
\begin{equation}\label{eq:query1}
\pi_{\{\pname,\city\}}(\suppliers \Join \parts \Join \projects \Join \LEQ).
\end{equation}

For the relations to be joinable, the signatures
$\tau_0$,
$\tau_1$,
$\tau_2$, and
$\tau_3$ have to be summable.
That is, any elements common to their source sets
have to have the same value.
For example, the source sets of
$\tau_0$ and
$\tau_1$ have \sid\
in common.
They both map \sid\ to its domain,
which is the set of supplier IDs.
Therefore
$\tau_0$ and
$\tau_1$ are summable;
hence $\suppliers \Join \parts$ is defined (Definition~\ref{def:join}).
Similarly with the other joins in the expression (\ref{eq:query1}),
which has as value the relation described by the \sql\ query in
Figure~\ref{fig:citiesParts}.

\begin{figure}
\begin{verbatim}
SELECT PNAME, CITY
FROM SUPPLIERS, PARTS, PROJECTS
WHERE PARTS.PID = PROJECTS.PID
      AND SUPPLIERS.SID = PARTS.SID
      AND RQTY <= PQTY
\end{verbatim}
\caption{\label{fig:SQL}
An \sql\ query to the relations in Figure~\ref{fig:citiesParts}.
}
\end{figure}

The query in Example~\ref{ex:citiesParts}
uses relations as they are are given.
Let us now consider queries joining relations
\emph{derived from} given relations.
Such derivations can be effected with filtering.

\subsection{Queries requiring filtering}

\begin{example}
\label{ex:parentChild}
In Figure~\ref{fig:parentChild} we show a table
specifying a relation consisting of tuples of two components
in which one is a parent of the other. 
It is required to identify pairs of persons
who are in the grandparent relation.
\end{example}

\begin{figure}
\begin{center}
$\pc=$
\begin{tabular}{l|l}
\parent & \child \\
\hline
\mary   & \john    \\
\john   & \alan     \\
\mary   & \joan    \\
\end{tabular}
\end{center}
\caption{\label{fig:parentChild}
Relation for Example~\ref{ex:parentChild}.
}
\end{figure}

What distinguishes this query
from the one in Example~\ref{ex:citiesParts} is
that the relations do not occur in the join as given,
but are derived from the given relation.
On the basis of the derived relations
we create one in which the pairs
are in the grandparent relation.
In one of the \sql\ dialects this would be:
\begin{verbatim}
SELECT PC0.PARENT, PC1.CHILD
FROM   PC AS PC0, PC AS PC1
WHERE  PC0.CHILD = PC1.PARENT
\end{verbatim}
In this query,
the derived tables are obtained via the linguistic device
of first renaming \pc\ to $\pc_0$ and then to $\pc_1$.
The \sql\ compiler translates such a query behind the scenes
to suitable relational operations.

Let us now return to Example~\ref{ex:parentChild}
to see how filtering is used here.
Suppose we have a set
$\Sym = \{x,y,z \}$
and an index set
$I = \{\parent,\child \}$.
Let $p$ and $q$ both be in $I \rightarrow \Sym$,
$p = \begin{tabular}{c|c}
$\parent$ & $\child$ \\
\hline
$x$ & $y$
\end{tabular}$,
and
$q = \begin{tabular}{c|c}
$\parent$ & $\child$ \\
\hline
$y$ & $z$
\end{tabular}$.

\paragraph{}
The relation
$$
\pi_{\{x,y\}}(\pc:p \;\Join\; \pc:q)
$$
is the equivalent of the relation resulting from the \sql\ query.
Here $\pc:p$ and $\pc:q$
denote the results of filtering $\pc$ with respect to $p$ and $q$,
respectively.

In this example, we have followed database usage in
making the index set
$I = \{\parent,\child \}$ non-numerical.
If we set $I = \iota(2)$,
then we can write
$[x,y]$ for $p$
and
$[y,z]$ for $q$.
The query then becomes
\begin{equation} \label{eq:parentQuery}
\pi_{\{x,z\}}(\pc:[x,y]\;\Join\; \pc:[y,z]),
\end{equation}
This bears a resemblance to the formula of
predicate calculus
\begin{equation}
\exists y.\; pc(x,y) \wedge pc(y,z).
\end{equation}
In the next section we will use the \etr\ operations
for semantics of predicate calculus.

In this paper we do not consider any query language;
we only discuss relational operations
and how they can be used to obtain
the computed relations that we want as results
from queries, independently of how they might be formulated
in a query language.

\begin{example}
\label{ex:square}
Consider the following instance of Figure~\ref{fig:network}.
$I = \iota(2)$,
$X = \{x,z\}$,
and
$T = \{R\}$.
Because $T$ has a single element,
there is only one possibility for $\tau$ and $\varphi$,
while $\sigma$ is pre-ordained by $T$.
Let us define
$$
prod =
\pair{\tau}{M}
\mbox{ with } 
M = \{\triple{u}{v}{w} \in \Rea^3 \mid u \times v = w\}
$$
and
$$
sq =
\pair{\varphi}{S}
\mbox{ with } 
S = \{s \in \{x,z\} \rightarrow \Rea \mid s_x^2 = s_z\}
$$
$prod$ is the ternary product relation;
$sq$ is the binary squaring relation.
In Figure~\ref{fig:network},
$t$ is a tuple in $M$ and 
$s$ is a tuple in $S$.

The fact that, and the way in which, squaring is a special case of
multiplication can be expressed as
$$
sq = prod:[x,x,z]
$$
\end{example}

\section{ETR as semantics for predicate calculus}
\label{sec:logSem}

Let us recapitulate Codd's plan for databases:
\begin{enumerate}
\item
data in the form of relations (the base relations), and
\item
queries, formulated in predicate calculus,
that define answers in the form of relations
and that are translated to machine-executable operations
on base relations
\end{enumerate}

As for (1),
we argued that Codd's understanding
of the mathematics of relations was not adequate;
we provided a theory
that answers to the needs of databases.
As for (2), again Codd gave a good start.
Unfortunately, it seems that the necessary follow-up
has failed to appear.
We address part (2) of Codd's plan in this section.

The attraction of predicate calculus
is that it is a formalism which is close to natural language
and has a mathematically defined semantics.
Early in the history of predicate calculus
this semantics existed only in the form
of an inference system.
A formula was considered true or false
according to whether the corresponding
truth value was derivable in the inference system.
This was valuable because the inference rules were
all deemed sound.
But could they be \emph{proved} to be so?
And could they prove everything that could reasonably
considered to be true?
These questions, the soundness and the completeness
of the inference system, could not be answered without
a \emph{declarative} semantics of logic.
Here ``declarative'' means independence
of any procedure, such as an inference system.
It is this declarativity that makes
predicate calculus potentially valuable for database:
we want a correctness criterion
for an implementation of the query language
that does not refer to some other mechanism.

In the 1930s Tarski proposed
a declarative semantics for predicate calculus.
The proposal was successful
because it corresponded to intuition
and because the intuition-inspired
  inference systems of the day
turned out to be sound and complete with respect to it.
Thus these inference systems
constituted \emph{procedural semantics},
as computer scientists were to call it later.
It must have been the fact that predicate calculus
has both a declarative and a procedural semantics
that recommended this formalism to Codd.

Codd's idea can be summarized as:
predicate calculus as a declarative query language
for the user
with an implementation
based on a sound and complete inference system.
At the time this was proposed
and implemented by Green \cite{grn69}.

However, the inference systems for predicate calculus
are not suitable for implementation of information
retrieval on the scale of databases.
If the difficulty needs to be summarized
in a few words one could say
that the inference system operates on a tuple at a time.
At the scale of a database one needs to operate on
meanings of predicate symbols (relations)
as a whole.

This suggests a third semantics for predicate calculus:
in terms of relations and operations on them.
That is, in terms of a relational algebra.
The pioneers in this direction were Halmos \cite{hlm62}
and, again, Tarski \cite{trsk52,trskth52}.
The applicability Tarski's relational algebra
to databases was noted by Imielinski and Lipski \cite{imlip84},
but this does not seem to have had an impact in the database world.

Codd saw the need for a relational algebra
as counterpart for a declarative query language.
He proposed the relational calculus,
a drastic modification of predicate calculus.
He defined a relational algebra corresponding to it.
In this way he obtained a mathematical basis for the
implementation of databases.
Because of the tenuous connection between relational
calculus and predicate calculus,
the declarative nature of the query language has been lost.
Thus it was left to others to restore
to databases the dual semantics that makes predicate calculus
so valuable:
a declarative semantics with an operational semantics
that is sound and complete.

So far we have only used \etr\ for a mathematical
definition of relations and operations on them.
We proceed to show how \etr\ can be used to provide
an operational semantics for predicate calculus.
What makes our semantics valuable for databases
is that it is defined in terms
of the operations on relations as defined in \etr.
At the same time it applies 
(see Theorems~\ref{thm:atom}, \ref{thm:compSem},
\ref{thm:exQuant}, \ref{thm:negation})
to predicate calculus
of which the declarative semantics has been established
equivalent to the proof theory of logic in the 1930s.
An alternative route would have been to show that
the declarative semantics of predicate calculus
can be adapted to Codd's relational calculus.
We have declined to take this route;
neither has it, to our knowledge, been taken by others.

A disadvantage for database use of predicate calculus
is that its tuples are restricted to numerical indexes
and that there is a single domain.
In our use of \etr\ as algebraic counterpart of
predicate calculus we restrict ourselves to this special case.
But we note that versions of predicate calculus have been
developed to accommodate multiple domains,
though still restricted to numerical indexes;
see ``many-sorted logic'' \cite{ndr72,gllr86}.
\etr\ has full generality of indexing
as well as many-sortedness.
To apply it, as we do here,
to the semantics of conventional predicate calculus,
does not use the full capability of \etr.

\paragraph{Satisfaction semantics of predicate calculus}

Conventional semantics is primarily concerned
with the justification of inference systems.
The use of predicate calculus
for the definition of functions or relations is secondary,
if considered at all.
As a result, conventional semantics
centres around the concept of \emph{satisfaction}:
under what conditions is a formula satisfied
by a given interpretation of the predicate symbols
and constant symbols under a given assignment
of domain elements to the variables.

For the sake of brevity
we assume the absence of function symbols,
in harmony with their absence
in conventional relational database theory.
Our language of logical formulas is determined by
a set $P$ of \emph{predicate symbols},
a set $V$ of \emph{variables}, and a set of \emph{constant symbols}.

\paragraph{}
A \emph{term} is a variable or a constant symbol.

\paragraph{}
A \emph{formula} can be
\begin{itemize}
\item
an \emph{atom} (or \emph{atomic formula})
is an expression of the form
$p(t_0,\ldots,t_{k-1})$,
where $p \in P$ is a predicate symbol and
$t_0,\ldots,t_{k-1}$ are terms.
\item
a \emph{conjunction} is
an expression of the form
$F_0 \wedge \cdots \wedge F_{n-1}$
where $F_i$ is a formula, $i \in \{0,\ldots, n-1\}$.
\item
an \emph{existential quantification} is 
an expression of the form $\exists x. F$
where $x$ is a variable and $F$ is a formula.
\item
a \emph{negation} is an expression of the form
$\neg F$, where $F$ is a formula.
\end{itemize}

An \emph{interpretation} for the language
consists of a set \D\ called the \emph{domain}
of the interpretation, a function \Pred\ that
maps every $k$-ary predicate symbol in $P$ to a subset of $\D^k$,
and a function \F\ that maps every constant symbol to an element
of \D.

An interpretation assigns the meaning $M(t)$
to a term $t$
and determines whether a variable-free formula is true
(in that interpretation). 
We refer to $M$ as ``meaning function''
or as ``interpretation''.
\begin{itemize}
\item
$M(t) = \F(t)$  if $t$ is a constant symbol
\item
A variable-free atom
$p(c_0,\ldots,c_{k-1})$
is satisfied by an interpretation iff
$ [M(c_0),\ldots,M(c_{k-1})]
   \in \Pred(p).
$
\item
A \emph{conjunction}
$A_0 \wedge \cdots \wedge A_{n-1}$
of variable-free atoms
is satisfied by $M$ if
$A_i$ is satisfied by $M$, $i \in \{0,\ldots, n-1\}$.
\item
A variable-free formula $\neg F$ is satisfied by $M$
if $F$ is not satisfied by $M$. 
\end{itemize}

We now consider meanings of formulas that contain variables.
Let \A\ be an \emph{assignment},
which is a function in $V \rightarrow \D$,
assigning an individual in $\D$ to every variable.
In other words, \A\ is a tuple of elements of $\D$
indexed by $V$.
As meanings of terms with free variables
depend on \A\, we write $M_\A$
for the function mapping a term to a domain element.
$M_\A$ is defined as follows.

\begin{itemize}
\item
$M_\A(t) = \A(t)$ if $t$ is a variable.
\item
$M_\A(c) = M(c)$ if $c$ is a constant symbol.
\end{itemize}
\begin{itemize}
\item
$p(t_0,\ldots,t_{k-1})$
is satisfied by $M$ and \A\ iff
$ [M_\A(t_0),\ldots,M_\A(t_{k-1})]
   \in \Pred(p).
$

\paragraph{}
Now that satisfaction of atoms is defined,
we can continue with:
\item
$F_0 \wedge \cdots \wedge F_{n-1}$
is satisfied iff the formulas $F_i$
are satisfied, for $i=0,\ldots,n-1$.
\item
If $F$ is a formula,
then $\exists x. F$ is satisfied by $M$ and \A\
iff there is a $d \in \D$ such that $F$ is satisfied with
$M$ and $\A_{x|d}$
where $\A_{x|d}$ is an assignment
that maps $x$ to $d$ and maps the other variables
according to \A.
\item
$\neg F$ is satisfied iff
formula $F$ is not satisfied.
\end{itemize}

If $S$ is a formula without free variables
(a \emph{sentence}) that is satisfied by no $M$,
then $S$ is said to be \emph{unsatisfiable};
if $S$ is satisfied by all $M$,
then it is said to be \emph{valid}.

\paragraph{Denotation semantics}

So far satisfaction semantics,
which is standard in treatments of predicate logic.
The purpose of these treatments is to
establish results like the soundness and completeness 
of inference systems
or to characterize the nature of logical implication.
The assigning of meanings to formulas with free variables
plays a subordinate role
in standard treatments of predicate logic.

One of Codd's requirements was that formulas with free
variables (used to represent queries)
be given as meaning a relation (the set of answers). 
This can be done with a generalization
beyond the satisfaction semantics summarized above. 
This generalization comes down to a generalization
of the mapping $M$ that extends its applicability
from closed formulas to open ones.
We call this generalization \emph{denotation semantics}.

Let us suppose that $F$ is a formula
and that $X$ is the set of its free variables.
We now consider
$$\{ t \in  (X \rightarrow \D) \mid 
            \exists \A. \; t = \A\downarrow X \mbox{ and }
   F \mbox{ is satisfied by } M \mbox{ and } \A\
\}.
$$

As \A\ is a tuple in $V \rightarrow D$,
$\A \downarrow X$ is a tuple in $X \rightarrow D$.
Hence this set is the extent of a relation with
$X \rightarrow D$ as signature.
This set of tuples is independent of \A,
so it is entirely determined by $F$ and $M$.
Accordingly, it can be used in the definition of
a meaning function \M\ to be applicable to
open formulas $F$, as follows.
\begin{definition}\label{def:openFormula}
Let $F$ be a formula with $X$ as set of free variables.
$$ \M(F) = 
  \langle X \rightarrow \D, 
                \{ t \mid \exists \A. \;
                t = \A\downarrow X \mbox{ and }
                F \mbox{ is satisfied by } M \mbox{ and } \A\
                \} \rangle.
$$
\end{definition}

Note that this is a relation with tuples
indexed by variables;
the relations that are the meanings of predicate symbols
are indexed numerically,
as forced by the syntax of predicate calculus.

Definition~\ref{def:openFormula} is a generalization of $M$
because it applies to closed formulas as well.
If $X$ is empty (we have a closed formula $F$),
then $\M(F)$ is a set of 0-tuples. As there is only
one 0-tuple, there are only two sets of 0-tuples.
Apparently a closed formula $F$ is unsatisfiable iff
$\M(F) = \{\}$.

\paragraph{An example from Codd}
As an application of Definition~\ref{def:openFormula}
we consider
Codd's \cite{codd70} (page 383)
definition of the ``natural'' join
between binary integer-indexed relations $R$ and $S$ as
\begin{equation}\label{eq:relComp}
R * S = \{ (a,b,c) : R(a,b) \wedge S(b,c)\}.
\end{equation}
This demonstrates a common confusion.
The expression $(a,b,c)$ refers to a triple of
elements of the domain; therefore
$a$, $b$, and $c$ are metalanguage names for
semantic entities.
By its form $R(a,b) \wedge S(b,c)$ suggests a formula
of logic, a syntactic entity.
Here 
$a$, $b$, and $c$ refer to variables of the formal logic.

It is not difficult to explain how this confusion came about.
What is meant by Equation~\ref{eq:relComp}
is the informal mathematical statement
\begin{quote}
``$R*S$ is the set of triples $(a,b,c)$
such that $(a,b)$ in $R$ and
$(b,c)$ in $S$.''
\end{quote}
There is no formal logic here:
$R$ and $S$ are names of relations,
$a$,
$b$, and
$c$ are names of domain elements.
Such a statement is often abbreviated to a shorthand such as
\begin{equation}\label{eq:relCompUS}
R * S = \{(a,b,c) : (a,b) \in R \mbox{ and } (b,c) \in S\},
\end{equation}
which is not confusing, because it is still clear that it is
entirely informal mathematics.

Usually such fine distinctions as we just made here do not matter.
However, when embarking on a project like the relational calculus
the distinctions become crucial.
We demonstrate the denotation semantics
developed in this section
by its definition of the ``natural'' join
between two relations.
Let $P$ contain the predicate symbols $r$ and $s$.
For the variables, let $V$ be $\{x,y,z, \ldots\}$.
With this syntax we can make formulas such as
$r(x,y) \wedge s(y,z)$.

On the semantic side,
we'll let domain \D\ be $\{a, b, c\}$
and consider the binary integer-indexed relations
$$\rho = \pair{\iota(2) \rightarrow \D}{\{[a,c],[c,b],[b,a],[b,b]\}}$$
and
$$\sigma = \pair{\iota(2) \rightarrow \D}{\{[a,b], [b,c],[c,a]\}}.$$

Assignments map variables to domain elements,
so \A\ could be a mapping with
$x \mapsto b$,
$y \mapsto a$, and
$z \mapsto b$.
It doesn't matter how \A\ maps
the other infinitely many variables.

Let's assume that the interpretation $M$ maps
$r$ to $\rho$ and $s$ to $\sigma$.
With this interpretation we
define a binary operation
$*$ on binary integer-indexed relations as
\begin{eqnarray*}
& &\rho * \sigma \\
&=&
\langle  \{x,y,z\} \rightarrow \D, \{t \mid \exists \A.\;
  t = \A \downarrow \{x,y,z\} \mbox{ and } 
  r(x,y) \wedge s(y,z) \mbox{ is satisfied by } M \mbox{ and } \A \} 
\rangle :[x,y,z]^{-1} \\
&=&  \pair{\iota(3) \rightarrow \D}
  { \{
[a,c,a],
[c,b,c],
[b,a,b],
[b,b,c]
  \} }
\end{eqnarray*}

From Definition~\ref{def:openFormula} it is clear
that, by taking the place of the $F$ in that definition,
only $r(x,y) \wedge s(y,z)$ is part of the formal
language of predicate calculus.

\paragraph{Properties of Definition~\ref{def:openFormula}}

Our claim that Definition~\ref{def:openFormula}
provides a semantics for the open formulas of
predicate calculus is based on the following theorems.
In these theorems we assume that the interpretation $M$
has \D\ as domain.

\begin{theorem}\label{thm:atom}
Consider the 
atomic formula
$ q(p_0,\ldots,p_{n-1}).$
Assume that $X$
is the set of variables in
$[p_0,\ldots,p_{n-1}]$.
Then we have
$$
\M(q(p_0,\ldots,p_{n-1})) =
\pair{
X \rightarrow \D
}{
M(q) : [p_0,\ldots,p_{n-1}]
}.
$$
\end{theorem}
\emph{Proof.}
According to Definition~\ref{def:openFormula}
the left-hand side has $X \rightarrow \D$ as signature.
So both sides have the same signature.
It remains to prove that the extents of the relations
are equal.

\paragraph{}
Let \A\ be such that $a = \A \downarrow X$. 

\begin{tabbing}
for margin \= $q(p_0,\ldots,p_{k-1})$ is satisfied by $M$ with \A\
\=$\Leftrightarrow$ (use satisfaction) \kill
\>$a$ in the extent of the left-hand side
\> $\Leftrightarrow$ (Definition~\ref{def:openFormula})\\
\>$q(p_0,\ldots,p_{k-1})$ is satisfied by $M$ with \A\
\> $\Leftrightarrow$ (use satisfaction) \\
\>$[a(p_0),\ldots,a(p_{k-1})] \in M(q)$
\> $\Leftrightarrow$ (use $t = a \circ p$) \\
\>$[t_0,\ldots,t_{k-1}] \in M(q)$
\> $\Leftrightarrow$ (definition of :)\\
\>$a \in$ extent of $M(q):[p_0,\ldots,p_{k-1}]$.
\end{tabbing}

\begin{theorem}\label{thm:compSem}
Let $X$ be the set of variables in the conjunction
$B_0 \wedge \cdots \wedge B_{n-1}$
of atomic formulas.
Then we have
$$\M(B_0 \wedge \cdots \wedge B_{n-1}) =
\pair
{X \rightarrow \D
}{
\M(B_0) \Join \cdots \Join \M(B_{n-1})
}
$$
\end{theorem}

\emph{Proof}\\
Both sides have the same signature.
It remains to prove that the extents are equal.

\paragraph{}
$a \in X \rightarrow D$ in the extent of the left-hand side\\
$\Leftrightarrow$ (Definition~\ref{def:openFormula})\\
$B_0 \wedge \cdots \wedge B_{n-1}$
satisfied by $M$ and \A\ such that $a=\A\downarrow X$,
where $X$ is the set of variables of
$B_0 \wedge \cdots \wedge B_{n-1}$\\
$\Leftrightarrow$ (definition of $\wedge$)\\
$B_i$ satisfied by $M$ and \A\ such that $a_i=\A\downarrow X_i$,
where $X_i$ is the set of variables of $B_i$,
for $i=0,\ldots,n-1$\\
$\Leftrightarrow$ (Definition~\ref{def:openFormula})\\
$a_i$ in extent of $\M(B_i)$
for $i=0,\ldots,n-1$
and signature of $\M(B_i)$
is $\tau_i = X_i \rightarrow D$
and $\tau_0,\ldots,\tau_{n-1}$
are summable\\
$\Leftrightarrow$ \\
$a$ in extent of
$\pi_{X_0}^{-1} \M(B_0)\cap \cdots
     \cap \pi_{X_{n-1}}^{-1} \M(B_{n-1})$\\
$\Leftrightarrow$ (definition of join)\\
$a$ in extent of
$\M(B_0)\Join \cdots \Join \M(B_{n-1}).$

\begin{theorem}\label{thm:exQuant}
Let $F$ be a formula with $X$ as its set of free variables,
and $Y= \{y_0,\ldots, y_{k-1}\}$ a subset of $X$.
Then we have
$$
\M(\exists y_0 \ldots y_{k-1}. F) = 
\pair
{(X \backslash Y) \rightarrow \D}
{\pi_{X\backslash Y}(\M(F))}
$$
\end{theorem}
\emph{Proof}
Both sides have the same signature.
It remains to prove that the extents are equal.

We take
$\exists y_0 \ldots y_{k-1}. F$
to be a shorthand for
$\exists y_0.\exists y_1. \ldots F$.

\noindent
$a \in X \backslash \set{y} \to D$
in the extent of the left-hand side\\
$\lra$ (Definition~\ref{def:openFormula})\\
$\exists y.F$ is satisfied by $M$ and \A\ such that
$\A \downarrow X \backslash \set{y} = a$\\
$\lra$ (definition of satisfaction, $\exists$ case)\\
$\exists d \in D$ such that $F$ is satisfied by $M$ and \A\
such $\A \downarrow X = a^d$
where $a^d$ such that $a^d(y) = d$
and $a^d(v) = a(v)$ for all $v \in V$ where $v$ is not $y$\\
$\lra$ (definition of \M)\\
$a^d$ in extent of $\M(F)$\\
$\lra$ (definition of $\pi$)\\
$a$ in extent of $\pi_{X \backslash \set{y}}\M(F)$.

\begin{theorem}\label{thm:negation}
Let $F$ be a formula.
Then we have
$$
\M(\neg F) = \M(F)^C
$$
\end{theorem}
where the complementation of the relation $\M(F)$
is according to definition~\ref{def:boolean}.

\emph{Proof.}\\
Both sides have the same signature.
It remains to prove that the extents are equal.

\begin{tabbing}
margin\=$F$ not satisfied by $M$ and \A\
  such that $\A \downarrow X = a$
\=$\lra$ (Definition~\ref{def:openFormula})\kill
\>$a \in X \to D$ in extent of left-hand side
\>$\lra$ (Definition~\ref{def:openFormula})\\
\>$\neg F$ satisfied by $M$ and \A\
  such that $\A \downarrow X = a$
\>$\lra$ (definition of $\neg$)\\
\>$F$ not satisfied by $M$ and \A\
  such that $\A \downarrow X = a$
\>$\lra$ (Definition~\ref{def:openFormula})\\
\>$a$ not in extent of $\M(F)$
\>$\lra$ (definition of complementation)\\
\>$a \in \M(F)^C.$
\end{tabbing}

\section{Related work}

Hall, Hitchcock, and Todd \cite{hahito75}
noted the confusion of Codd and subsequent authors
resulting in ``relations'' as distinct from ``relationships''.
They defined relations with non-numerical indexes
as the correct remedy.

The semantics given here applies only to first-order predicate
calculus without function symbols.
This restriction is lifted in \cite{vnmdn06}.
But there the semantics applies
not to the classical syntax of predicate calculus,
as presented here,
but to the Horn-clause subset of the clausal syntax.

Clark \cite{clrk91} seems to be the first to explicitly
define a denotation semantics for atomic formulas.

\section{Conclusions}

We see three main contributions of this paper.
The first is to develop \etr, an elementary theory
of relations adequate for expressing relational
databases, queries, and answers.
This theory is entirely in terms of elementary set theory,
the same set theory that underlies all mathematics.
\etr\ should be compared with Codd's relational algebra,
created solely for the purpose of supporting relational calculus.

The second contribution is to show that the special-purpose
relational calculus is not necessary.
We show that queries can be formulated in predicate calculus,
the same logical calculus that is the formal logic preferred
for the formalization of mathematical theories since well before
the advent of databases.

The third contribution arises as a by-product of the first two.
Its significance is independent of database theory:
a generalization of the traditional satisfaction semantics
of predicate calculus.
Denotation semantics, as we call this generalization,
is suited to applications elsewhere in computer science,
as it makes predicate calculus into a powerful and flexible
tool for defining new relations in terms of existing ones.
The traditional satisfaction semantics
remains of course satisfactory
for the traditional purpose of formal logic:
to analyse formalized theories for validity
and to investigate the soundness and completeness of
proof procedures.

\section{Acknowledgements}
We benefited from discussions with Areski Nait-Abdallah.
We thank Victor Marek for pointing out errors.
The research was supported by the University of Victoria and by
the Natural Science and Engineering Research Council of Canada.

\end{document}